# Point estimates, Simpson's paradox and nonergodicity in biological sciences

Madhur Mangalam and Damian G. Kelty-Stephen

Madhur Mangalam is in the Department of Physical Therapy, Movement and Rehabilitation Sciences, Northeastern University, Boston, MA, USA; Damian G. Kelty-Stephen is in the Department of Psychology, Grinnell College, Grinnell, IA, USA.

E-mails: m.manglam@northeastern.edu; keltysda@grinnell.eud

## Abstract

Modern biomedical, behavioral and psychological inference about cause-effect relationships respects an ergodic assumption, that is, that mean response of representative samples allow predictions about individual members of those samples. Recent empirical evidence in all of the same fields indicates systematic violations of the ergodic assumption. Indeed, violation of ergodicity in biomedical, behavioral and psychological causes is precisely the inspiration behind our research inquiry. Here, we review the long term costs to scientific progress in these domains and a practical way forward. Specifically, we advocate the use of statistical measures that can themselves encode the degree and type of non-ergodicity in measurements. Taking such steps will lead to a paradigm shift, allowing researchers to investigate the nonstationary, far-from-equilibrium processes that characterize the creativity and emergence of biological and psychological behavior.

# BOX 1 GLOSSARY

**Effect size.** The magnitude of the observed effect of an independent variable on a dependent variable. The effect size obtained for a sample is an estimate of the population effect size.

**Gaussian distribution.** A frequency distribution that arises from additive processes, is well defined mathematically, and is assumed to be common in empirical data.

**Hurst exponent:** A measure of long-range temporal correlations in time series.

**P-value.** The probability of the same result or one more extreme just by chance if the null hypothesis is true. When dichotomized according to experimenter-determined tolerance for false-positive error, this value serves as an inverse measure of the strength of evidence against the null hypothesis.

**Population.** The set of all individuals of a specific type that is to be statistically measured but is too vast to be sampled exhaustively such that an exact measure of the population cannot be obtained.

**Regression.** A family of statistical techniques used to estimate the relationships between a dependent variable (often called "outcome") and one or more independent variables (often called "predictors").

**Replicate.** To repeat a procedure using a new sample from the same population(s).

**Reproduce.** To find the same results when a procedure is repeated using a new sample from the same population(s).

**(Random) sample.** Observations obtained randomly from a defined population used to estimate population characteristics.

**Sample size.** The number of observations in the sample.

**Statistical power.** A measure of the capacity of a statistical test to yield a significant result. It depends on the threshold for significance, size of the expected effect, variation in the population, the alternative hypothesis (one or two-sided), nature of the test (paired or unpaired), and sample size.

**Variance.** The spread between observations in a sample, quantifying how far each observation in the sample is from the mean, and therefore from every observation in the sample.

**Box 2 ERGODICITY**

A stochastic process $x(t)$ can be subjected to two types of averaging: the ensemble average and the time average. The finite-ensemble average of the quantity $x$ at a given time $t$ is

$$\langle x(t) \rangle_N = \frac{1}{N} \sum_i^N x_i(t) \tag{1}$$

where $x_i$ is the $i$th realization of $x(t)$ and $N$ is the number of realizations included in the average. The finite-time average of the quantity $x(t)$ is

$$\overline{x_{\Delta t}} = \frac{1}{\Delta t} \int_t^{t+\Delta t} x(s)\,ds. \tag{2}$$

If $x$ changes at $T = \Delta t / \delta t$ discrete times $t + \delta t,\ t + 2\delta t, \ldots$ , then Eq. (2) becomes

$$\overline{x_{\Delta t}} = \frac{1}{T\,\delta t} \sum_{\tau+1}^T x(t + \tau\,\delta t). \tag{3}$$

An observable $X$ is ergodic if its ensemble average converges to its time average with probability one, such that

$$\lim_{N \to \infty} \frac{1}{N} \sum_i^N X_i(t) = \lim_{\Delta t \to \infty} \frac{1}{\Delta t} \int_t^{t+\Delta t} X(s)\,ds. \tag{4}$$

Hence, a stochastic process is ergodic if any random collection of samples represents the entire process's average statistical properties. Conversely, a stochastic process is nonergodic when the statistical properties of that process change with time. An unbiased random walk is nonergodic, as its time average is a random variable with divergent variance about the expectation value of zero. For instance, let us choose at random one of the two fair coins and then toss the selected coin $n$ times. Let the outcome be 1 for heads and 0 for tails. Then the ensemble average is ½(½ + ½) = ½, which is equal to the long-term average of ½ for either coin. Hence, this random process is ergodic. Conversely, let us choose at random one of the two coins: one fair and the other has two heads, and then toss the selected coin n times. The ensemble average for this case is ½(½ + 1) = ¾; yet the long-term average is ½ for the fair coin and 1 for the two-headed coin. Hence, this random process is not ergodic.

Note that ergodicity is distinct from stationarity. An example of a stationary but nonergodic process is rolling a dice today. The function $x(t)$ for all future dates is whatever is rolled today. $x(t)$ is clearly nonergodic given that its expectation value is 3.5, but its time average is either 1, 2, 3, 4, 5 or 6. Nonetheless, it is stationary. Meanwhile, it is possible to have an ergodic process that is nonstationary, precisely when the expectation value is static but the variance. For instance, during a jazzy drum solo, the high-hat cymbal is locked into position on the drum kit, but the intermittent strikes by the drummer will dramatically change the size of the oscillations in the cymbal from beat to beat and measure to measure.

**Box 3 SIMPSON'S PARADOX**

Simpson's paradox is a phenomenon in which a statistical trend observed at the individual level (or at the level of groups) either disappears or reverses at the group level (or at the level of a set of groups). The paradox can be resolved when individual relationships are appropriately addressed in the statistical modeling. Simpson's paradox is especially problematic in biomedical sciences, where trends observed at the group level are often fallaciously used to derive inferences about individuals. An often-quoted intuitive example of Simpson's paradox is the correlation between typing speed and typos[1]. At the group level, the correlation is negative—more experienced typists type faster as well as make fewer typos (solid black line in Fig. I). However, at the individual level, the correlation is positive—the faster an individual types, the greater the number of typos he/she makes (blue dotted lines in Fig. I). Thus, it would be fallacious to conclude that the relationship between typing speed and typos observed at the group level holds for individuals. The interested readers can refer to some of the excellent resources that discuss statistical methods to prevent, diagnose and treat Simpson's paradox in point data[2–4].

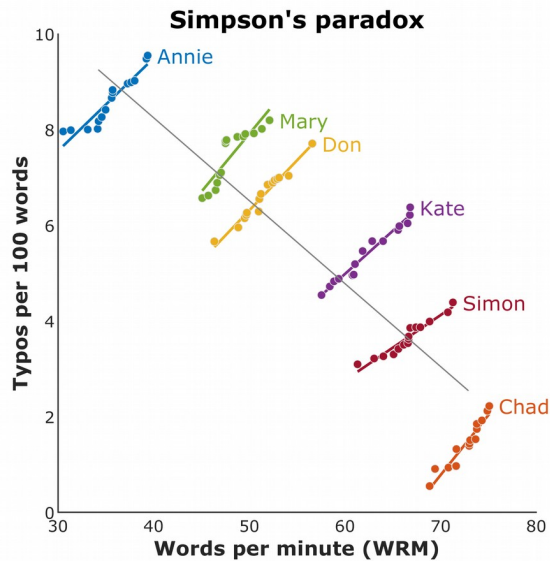

**Fig. I.** Relationship between typing speed and typos. Simulated data illustrating that despite a negative correlation at the group level, within each individual, there exists a positive relationship between typing speed and the frequency of typos.

**Box 4 STATISTICS FOR QUANTIFYING NONERGODICITY**

Ergodicity refers to the resemblance of sample variance to population variance. Statistical methods often phrase this question as testing how much subsample variances vary from the bigger sample's variance, that is, the variance of sample variances. This resemblance can be considered in two ways: first, in terms of whether the time-average variance of any single trajectory resembles the ensemble-average variance of those multiple trajectories and, second, in terms of whether the time-average variance of a subset within a single trajectory resembles the time-average variance for the entire single trajectory.

Consider a sample of participants, each flipping a coin repeatedly 200 times. If we assume that each coin is fair, and each participant performs the flips similarly, the variance in outcomes should presumably resemble each other in the long run. We can test this assumption by looking at any one participant's flipping, taking the variance for that participant's whole sequence, and comparing it with the variance across all participants. Deviations in that participant's flips might suggest a weighted coin or a systematically different flipping method. If that participant's flipping behavior does not resemble the average behavior across all participants, then ergodic assumptions fail. Of course, each participant is more likely to deviate from a larger subsample of participants. So, it is possible to examine variance across progressively larger subsamples of individuals (e.g., 1 participant alone, 2 participants, 4 participants, and so forth) and test how quickly subsample-average variance converges towards whole-sample-average variance. Now, we can also consider how well subsamples of any one participant's flipping outcomes resemble the variance across the whole sequence: because the small-sample bias likely makes variance unstable, we do not expect all subsamples of 5-flips sequences to yield the same estimated variance as the entire 200-flips sequence. However, we can evaluate the average variance for gradually longer subsets (i.e., for 5-, 10-, 20-flips sequences, and so forth) and evaluate ergodicity as the speed at which the variance across these subsets converges towards the variance across the whole sequence.

To our knowledge, the most precise way to portray ergodicity is through a dimensionless statistic of ergodicity breaking $E_B$—also known as the Thirumalai-Mountain (TM) metric[5]—that subtracts squared total-sample variance from average squared subsample variance and then divides by the total-sample squared variance (Fig. II).

$$E_B\big(x(t)\big) = \frac{\left(\left\langle\left[\delta^2\big(x(t)\big)\right]^2\right\rangle - \left\langle\overline{\delta^2}\big(x(t)\big)\right\rangle\right)}{\left\langle\overline{\delta^2}\big(x(t)\big)\right\rangle^2} \tag{5}$$

which is zero in the limit $t \to \infty$ if the process is ergodic. In effect, this statistic phrases deviation of squared subsample variance from squared total-sample variance as a proportion of squared total-sample variance. Evidence for ergodicity is the rapid convergence of this statistic to zero for progressively larger samples. Slower (or non-) convergence indicates weaker (or non-) ergodicity[6]. Here, we would like to caution about ergodicity breaking using an example from economics. Models of relative wealth (i.e., what proportion of total national wealth an individual owns) often assume ergodicity, that is, fast convergence to a stable asymptotic distribution. This assumption can be upheld against all evidence: it gives answers, puts numbers on parameters. It is when we work outside this assumption that we find that it is unjustified[7]. Hence, sometimes we may not see a fast convergence to any asymptotic distribution, sometimes no convergence at all.

***

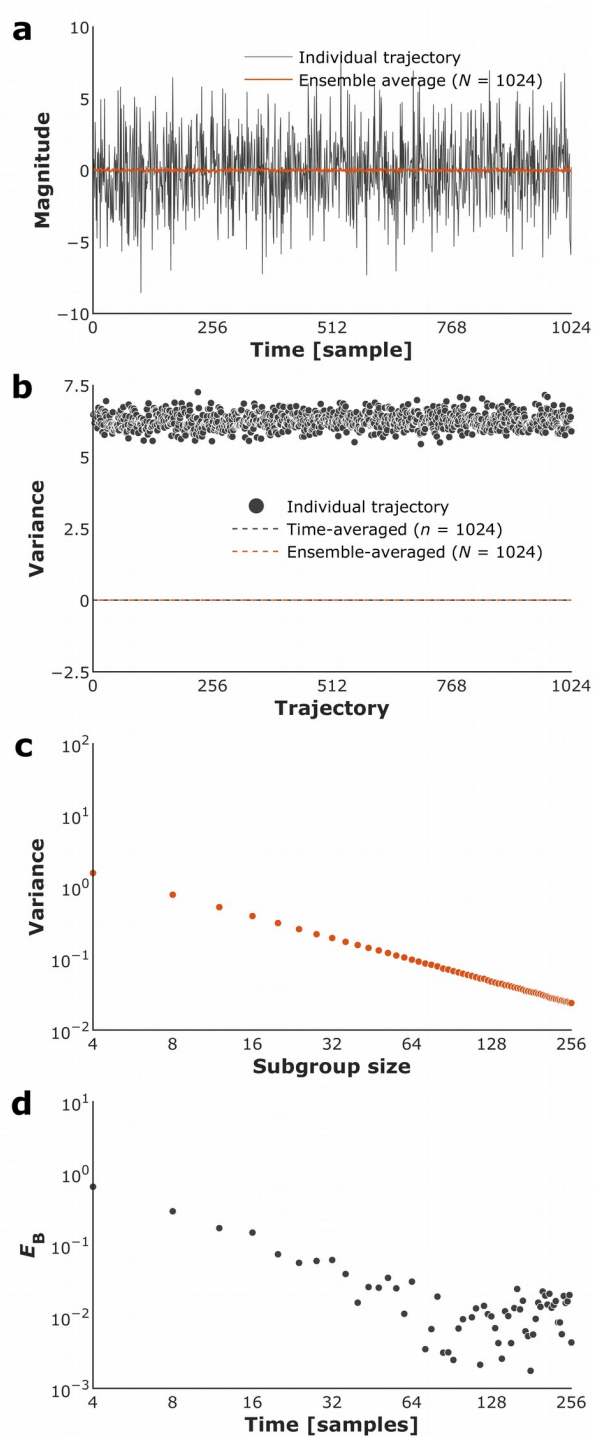

**Fig. II.** Ergodicity in additive white Gaussian noise (AWGN). (**a**) A representative trajectory of while noise and ensemble average of $N = 1024$ such trajectories. (**b**) Variances of individual trajectories, as well as time-averaged and ensemble-averaged variances across all 1024 trajectories (the latter two are the same for AWGN). (**c**) Ensemble-averaged variance calculated over differently-sized trajectory subgroups. (**d**) The ergodicity breaking parameter $E_B$ versus time $t$.

The universe is not ergodic.

—Sean Carroll

## Introduction

Scientific advancement depends on the reproducibility and validation of research findings. Poorly reproducible studies can impede and misdirect scientific progress, jeopardize funding, and lead to harmful clinical applications. Growing awareness among the scientific community about lapses of reproducibility in biomedical sciences[8–11], including psychological sciences[12,13] and neurosciences[14–17] have inspired recent developments like the Nature series entitled "Challenges in irreproducible research," the Reproducibility Initiative, a global project intended to identify and reward reproducible research (http://validation.scienceexchange.com/#/reproducibilityinitiative), and increased transparency and data sharing practices[18–20]. Although we loud these efforts, the focus has mostly been on the fallacies of P-values[21,22], small sample size[23,24], inaccurate estimation graphics[25,26], and reporting biases[27]. A fundamental problem pervasively linked to the lack of reproducibility in human subjects research—inherent in standard analytical techniques—that remains to be considered is nonergodicity, the paucity of group-to-individual generalizability[3,28–32]. Crucially, despite the expectation that group treatments will inform individual-level interventions or outcomes, human-subjects research may take ergodicity for granted when it should not. Here, we aim to address how recognizing and quantifying nonergodicity holds great promise for supporting future scientific progress.

Biomedical, behavioral and psychological studies infer from statistical tests conducted on aggregated data. This policy rests on the assumption that group-level statistical effects can be applied to understanding the physiology and psychology of an individual—that is, they assume that the study phenomenon is "ergodic"[28–30,33] (Box 2). Ergodicity holds when individual-level variability is homogeneous in resembling variability at the level of the group (or "ensemble" in statistical-physical parlance) and when individual-level variability is stationary, exhibiting homogeneous mean and variance over time[2,34,35]. Unfortunately, the differentiation of form and behavior inherent to biological structure violates the assumptions of homogeneity and stationarity, respectively, inevitably making it nonergodic. This violation is no peripheral nuisance but follows systematically from all of our most unambiguous attempts to distinguish life as an evolving, innovating, and adaptive process distinct from non-living processes[36]. The second condition, stationarity, rules out most physiological and psychological processes with time-varying moments (mean function, sequential covariance function) of being ergodic (e.g., heart rate variability, motor control, developmental processes, learning processes, and transient brain responses)[37]. Sample-level statistical modeling that assumes stationarity where it is not, cloaks artifacts of nonstationarity or fails to articulate systematic changes producing nonstationarity, obscures any genuine individual differences, and jettisons any generalizable truths we might have gleaned from the same diversity initially intended to represent the population. In these circumstances, deriving inferences from statistical tests conducted on aggregated data might profoundly threaten our goals of scientific consilience and completeness. It certainly makes the suspicious compromise of enforcing similarity and ignoring diversity for the sake of formal convenience.

Here we argue that behavioral, psychological, and neuronal processes unavoidably violate the ergodic assumptions, with reference to typically-studied phenomena. We discuss how, for this reason, a significant body of work in these fields falls short on scientific investigation's fundamental objectives, ranging from the failure to reproduce research findings to the inability to test hypotheses about nonlinear far-from-equilibrium dynamics. We explain why finding meaningful ergodic observables is essential for investigating nonergodic processes, and we introduce a few strategies that allow this transformation.

## Behavioral, physiological, and neuronal processes can violate ergodic assumptions

Despite multiple calls about the perils of violating ergodic assumptions[29,31,32,38,39], extant work in biomedical sciences has been mostly based on best-practice guidelines almost exclusively based on statistical inferences from data aggregated across large samples. Whether couched in prosaic terms or using formal mathematical theorems, the violation of ergodic assumptions questions the validity of this work. Here we provide a few examples of research areas that are particularly vulnerable to the violation of ergodic assumptions.

- Physiological sciences. One area in human physiology where nonergodicity has been shown to be particularly important is the biophysical transport of liquids. Such nonergodic processes include hemodynamics and intracellular and extracellular transport of complex media in biological systems, such as cytoplasm and nucleoplasm[40,41]. For instance, to reach every cell, the blood must enter every cell continuously in the temporal domain. Due to the fractal organization of the vascular network, the blood flow decelerates at small length scales (i.e., in finer capillaries). The blood cells spend more time at small volumes in comparison to the volumetric fraction of these volumes, making these systems nonergodic. Consequently, calculations using the time fraction, (corresponding to the time average) do not apply to calculations using the volumetric fraction (corresponding to the ensemble average)[42].

- Behavioral and psychological sciences. A well-known phenomenon from cognitive psychology in which the individual- and ensemble-level findings often conflict is the speed-accuracy trade-off[43]. Changes in the speed-accuracy tradeoff form the bedrock of motor learning[44]. Although speed and accuracy show negative correlation across individuals, individuals show a negative relationship between speed and accuracy, reflecting differential use of strategies to achieve the task goal[45]. Indeed, people often use multiple cognitive/motor strategies to achieve the same task goal, exhibiting an adaptive combination of accuracy and speed[46]. This phenomena is analogous to Simpson's paradox (Box 3), which is ubiquitous in behavioral and psychological sciences. In a classic study, Siegler[47] identified three factors that can lead to incorrect conclusions about data averaged over strategies: 1) Relative frequency of each strategy. The more frequently a strategy is used, the greater its influence on the average data. 2) Relative variability of performance from each strategy. Among strategies producing unequal variance on the dependent variable, the one with higher variance will have a larger influence on the average data. A less frequent strategy can have a greater influence if it leads to a much higher variance. 3) Variability in the relationship between independent and dependent variables within and across strategies. A high correlation between the smallest addend and the performance can falsely indicate that the strategy with the smallest addend was used more frequently than it was. Hence, psychological studies might overestimate the relationship between a set of variables at one level over other level(s) due to the clustering of strategies more often than commonly thought[2].

- Neurosciences. In neurophysiological studies, neuronal spikes show temporal fluctuations all along the stimulation period. The rate coding paradigm assumes that these fluctuations do not contain any information. So, this paradigm eliminates such statistical fluctuations by averaging over many trials. Indeed, the assumption is that the central nervous system is obliged to make an ensemble average over many neurons with the same function, that is, neuronal spiking is ergodic. However, arguments supporting ergodicity in spike trains are weak. For instance, one argument posits that approximately 30 neighboring neurons may form a functional unit, effectively encoding the same stimulus[48]. The central evidence supporting this conjecture is that neighboring neurons receiving shared inputs belong to the same functional unit[48–50]. However, because neurons require chain-reaction, they cannot resemble each other. Neurons also receive heterogeneous inputs from neighboring neurons and hence show nonergodicity. For instance, if in a population of neurons that show a synchronous peak in firing rate at the ensemble level, the timing of the peak drifts from trial to trial, then averaging the single-cell firing rate over trials would not show any peak. In contrast, in a coupled network of heterogeneous neurons, each neuron may show a peak at a different time that does not drift with trials; in this case, ensemble-average firing rate can be flat because population heterogeneity will mask the peaks for individual neurons. Hence, neurophysiological studies based on ensemble-level averaging of neuronal spiking data might violate the ergodic assumptions, portraying an erroneous picture of neuronal function at the single-cell level[51].

Thus, diverse mechanisms rooted in the geometry of the biological system (e.g., fractal structure of the vascular network, multiplicative interactions among cognitive processes, and heterogeneity in neuronal connectivity) lead to nonergodicity. Indeed, nonergodicity is more likely to be overlooked when studying such biomedical systems, particularly when obtaining long observations of time series is not feasible. Processes that involve some growth, such as development and aging, and are naturally susceptible to practical constraints of data collection, are particularly vulnerable to such oversight [29,32].

## Neglecting ergodic assumptions undermines the scientific enterprise

The widespread mistreatment of nonergodicity in biomedical and psychological sciences has substantial epistemic and practical consequences, ranging from none in the case of the few ergodic processes in nature, to catastrophic if a process is highly nonergodic. Indeed, neglecting ergodic assumptions might jeopardize the scientific community's very many efforts in addressing the challenges of P-values[21,22], small sample size[23,24], inaccurate estimation graphics[25,26], and reporting biases[27]. In many cases, these consequences set back the following four very objectives that form the bases of these research areas:

- Failure to replicate research findings. Failures to replicate research findings might be baked into the potentially incomplete, but broadly sweeping failure of human behavior and measurements of such class of systems to conform to the standard of independent and identically-distributed behavior around the mean value. In place of this assumption, to remove uncertainty in the measurement, the standard technique is average across time to extract the true value. Although this process yields a reliable, temporally stable value, this temporal stability is no guarantee of later reproducibility: performing the same experiment again will yield a different result every time. For instance, we will find that time-averaging Brownian motion over time will yield a progressively smoother trajectory. This result will lead us to believe that it has converged to some fundamentally true value, but repeating this exercise will yield a different value[52]. The ergodic hypothesis is especially apt in dealing with the processes that visit all possible states in a finite sample space. For instance, a good model of the numbers that show up in roulette wheel is is closely ergodic. The probability distribution of numbers (0 to 36) that have come up in the past is the same as the probability distribution of numbers on the next spin. In contrast, if a human participant is asked to say a number between 0 and 36, that participant might show systematic bias towards smaller or larger numbers. The bias might also depend on exogenous factors such as the context, surroundings, and time of the day[53,54]. Therefore, a good model of the numbers produced by humans is nonergodic, as the human behavior defies the ergodic assumption of independent and identically-distributed behavior around the mean value. Applying ergodic statistics to these nonergodic processes leads to wrong conclusions.

- <u>Failure to probe nonlinear dynamical principles underlying the study phenomena</u>. Ergodic statistics at best can describe the effects of manipulations on the outcome variable assuming ergodicity holds. They are not equipped to decipher the nonlinear dynamical principles underlying the study phenomenon. The ergodic assumption implies that a given manipulation for a participant in a study is always mediated by the same components, in the same manner, to link it with some measured output. An alternative view is interaction-dominant dynamics (IDD): component-causal chains are not the causal building blocks of behavior but are themselves self-assembled; that is, they emerge out of and do not exist independently from task constraints[55,56]. Ergodic statistics that contrast distinct groups of treatments with a between-participant design do not allow to trace IDD-related effects of time or influences of task changes that could lead to phase-transitions between conditions leading to different interpretations of the same treatment. Consequently, the assumption that consecutively measured values of behavior or physiology are interdependent violates the IID-assumption such that the variance of a measured observable cannot be parsed into specific, stable, and generalizable sources of variance reflecting the dynamical principles underlying the study phenomenon. Indeed, cognitive science has been explicit that when we instruct human participants to generate replicable behaviors in sequence, our instructions are in vain: "When the optimal strategy in a task is to provide a series of independent and identically distributed responses, people often perform suboptimally"[57]. So, for instance, if we leave human participants to count off seconds without any feedback from a clock, their behavior will be relatively non-ergodic, suggesting that their sense of how long a second lasts will narrow and dilate across time.

- <u>Failure to test hypotheses about nonlinear, far-from-equilibrium dynamics</u>. The ergodic assumption of equivalence between the ensemble average and the time average of an observable is a key component of equilibrium dynamics. When this assumption is valid, dynamical descriptions can be replaced with much simpler probabilistic ones, essentially eliminating time from the models. The conditions for validity are often restrictive for non-equilibrium dynamics and even more so for far-from-equilibrium dynamics. Biomedical sciences inevitably deal with far-from-equilibrium systems—specifically with models of growth and stability. For instance, standing quietly and maintaining focus on a target in front of us is the preamble to very many coordinated behaviors—we might lean forward and reach or track the target's progress and bat it away. However, this starting position is not merely the preamble to action but is already a rich wellspring of action itself, exhibiting a continuous stream of intermittent fluctuations. So long as they do not pitch the bodily center of pressure (CoP) beyond the base of support, these fluctuations are crucial to maintaining a quiet stance[58,59]. Therefore, it is surprising that the prevailing statistical methods make an indiscriminate assumption of ergodicity, feeding the bias to study systems while they are in equilibrium because only then can the behavior be measured to obtain a point estimate. But nonlinear dynamical systems reveal themselves when they are far from equilibrium, and this is when the dynamical principles come into play and nongodicity reigns. We can take the previous example of participants responding every time they think a second has passed. If we now give our second-estimating participants a clock to provide feedback on their accuracy, this task constraint can make behavior look significantly more ergodic[60]. Then again, it is when we pose the more open-ended questions, not just removing feedback but giving participants insight problems to prompt more creative response—then we see that the mind at its most creative looks dramatically less ergodic, with fits and starts, with distractions, and sudden "aha!" moments[61].

- Failure to articulate new hypotheses based on the current findings. A new theory may suggest its initial hypotheses. The scientific program depends on testing these hypotheses and generating new ones in the process. However, if the theory invokes nonlinear, far-from-equilibrium processes like emergence, then it presumes nonergodicity at its foundations. Statistical modeling that assumes ergodicity of raw measurements in these processes risks undermining such theorizing. If we use a linear model assuming ergodicity, we never test the original hypotheses about a nonergodic process. For instance, the emergent coalition model (ECM) of word learning and vocabulary development predicts that lexicon self-organizes from a coalition of exogenous cues[62]. Testing the effects of a set of cues in a linear model assuming ergodicity will always operate by estimating independent factors assumed to hold invariantly across time and individual participants. The trouble here is that linear modeling is a filter on our measurements. It presupposes ergodicity in the measurements entered into the model, and then it estimates sizes and directions of effects exclusively in ergodic terms. Any contingency of effects on time, space, or participant variability must be transparent to the model in terms of our preparation and coding of the data we feed into the model. For instance, an autoregressive moving-average model (ARMA) of measurement with trends over time will repeatedly fail to converge with stable residuals because the inputs are presumed to be stationary. It requires the scientist either to take the first difference of the data with trends before running the ARMA model or to change the modeling strategy to an ARIMA (including the "I" for "integration") that reflects a modeler's explicit coding of the linear trend with time. If the trends are fractional or nonlinear, ARIMA may or may not converge but will definitely fit the wrong trend and estimate effects for variables that may not exist and underspecify the complexity of the measurement. The linear filter will not know what nonergocities it may be missing, and so it will give no insight into the presumed non-equilibrium mechanisms driving development. Now, in the case of language learning, when a linear model finds that cue X (e.g., maternal pointing) predicts outcome Y (e.g., learning count nouns), then the linear filtering removes any nonlinearity, and suddenly the output looks as linear and ergodic as the theory allegedly predicted the phenomenon was not. Nothing about a linear modeling output on learning more count nouns from maternal pointing will then suggest as to which class of emergent phase shifts is in play, which populations will exhibit which individual differences in terms of which count nouns they learn or whether the count nouns they learn align with parallel syntax development. The ergodic assumptions suggest that all trajectories are the same on average, and all of these cues and subsets of language performance would be assumed parallel by any linear model. Of course, we could come up with new hypotheses about how far-from-equilibrium dynamics leads to a given emergent structure, but ergodic models cannot "see" the underlying nonlinearity and temporal variation. No matter the rhetoric about nonlinear processes like emergence, unless we are testing predictions that explicitly nonlinear sources of nonergodicity [e.g., as ARIMA does for linear sources[63]] the linear models and their independent and identically distributed estimates are themselves mute to nonlinear dynamical facts like emergence-not just unable to test the initial hypothesis about non-equilibrium dynamics like emergence but also unable to refine new hypotheses relevant to non-equilibrium dynamics.

Hence, all human subjects research that ignores the nonergodicity of study phenomenon falls short on these four fundamental objectives of scientific investigation. In basic research, we must not overstate how misleading our impressions might be about the effects of physiological/psychological manipulations and their interactions. In the clinical domain, diagnostic tests might be systematically biased, and our classification systems and treatments/interventions may be at least partially invalid. In medicine, this calls for personalization. For instance, a drug might not be effective unless he has some gene(s). By testing a large population, one would (correctly) conclude the drug is effective in 90% of the cases. It would be incorrect to conclude that an individual would respond to the drug 90% of the time it is tested on that one person—it may be yes/no at the individual level. In studying a new phenomenon, we might be not even employing the research designs necessary to adequately test the first line of hypotheses, let alone articulate new theories based on the current findings, as discussed above. We can research with the wrong glasses on, and spend our careers finding all sorts of effects and inventing nomenclatures without ever addressing a much more serious, all-encompassing problem. It can be a check-mate situation: irrespective of decades of research, we may have to throw it all away as invalid because it rests on an invalid ergodicity assumption. It is high time that we embrace nonergodicity and teach the next generation of scientists some nonergodic research designs and statistical techniques. Here, we side with Molenaar[29] that the implications of the classical ergodic theorems imply that focusing on finer-grained variation is not optional. The systematic violation of of these theorems in our raw measurements makes this focus a necessity.

## Multiscaled estimates of nonergodic measurements can behave ergodically

Effects that look nonlinear and not stable in linear modeling are often better behaved when the modeling allows causality to unfold across diverse scales. This point does not imply that an observed phenomenon is independent at different scales but that linear models can respect multiscaled texture. There are two ways this can happen and might be relevant. First, mixed-effects modeling allows estimating "fixed" effects for explicit manipulations spanning the whole task and "fixed" effects of explicit manipulations that vary across the task (e.g., blocks with and without treatment), as well as "random" effects capable of absorbing some rudimentary ways in which participants might differ in the intercept of their responses and in the slope of those responses across trials. That said, the references "fixed" and "random" are misnomers to the degree that very many explicitly randomized manipulations across the task (e.g., levels of an informational variable) can show the so-called "fixed" effects. For instance, a given level of an informational variable can show different effects in Block 1 versus Block 2, and that level could have been delivered on a randomly different trial within those two different blocks. However, the availability of the "fixed" effect of that informational variable means that this modeling allows fitting an estimate of the population-level effect of the experimenter-randomized informational variable. Taking this a step further, we can also test the "fixed" effects of endogenous variables that participants bring to each trial within each block, although not undermining the fact that participants bring individual differences to the task on each trial. It is only to say that some of the endogenous variety in how participants meet the task constraints could represent causal mechanisms generic to the whole population. For instance, the fact that different participants will direct different numbers of gaze fixations on a visual display will not prevent a mixed-effect model from estimating the population-level effect of gaze fixations on visual-perceptual response.

Again, the nomenclature: the random walk (process 3) is not ergodic, but its steps (process 1) are ergodic. Because its steps are ergodic, so is, for instance, its square deviation over some fixed interval. Finding meaningful ergodic observables for nonergodic processes is a hugely important part of doing ergodicity economics, and of doing science in general. This brings us to the second way we can make linear models respect multiscaled texture, which is by coming up with some form of statistics that quantify nonergodicity can themselves empower ergodicity-requiring models (e.g., vector autoregressive, VAR models) to model what would otherwise be too nonstationary to model. For instance, fractals in the nervous system: conceptual implications for theoretical neuroscience[64]. Submitting raw position/physiology/kinematic values into VAR models leads to poor convergence because the residuals are never serially independent. The failure to converge is because the raw values are serially dependent across time for reasons beyond linear correlations. Yes, there are vector autoregressive fractionally integrated (VARFI) vector autoregressive fractionally integrated moving-average (VARFIMA) models that can fit vector (i.e., multivariate) autoregression relationships around long-range fractional integration (the "FI" in "VARFI" and "VARFIMA") sort of memory[65]. But these solutions are limited: the VARFI and VARFIMA models are computationally expensive, and elegant methods like detrended cross-correlation analysis[66], can only treat with two variables and do not estimate unique effects of either.

On the other hand, VAR modeling of fractal and multifractal estimates yields stable residuals in accordance with the ergodic assumptions, and they have repeatedly shown both systematic responses to experimental manipulations and predictive relationships with individual participant responses[67,68]. Although some moderate failure to have serially independent residuals leaves fractal Gaussian noise (fGn) ergodic, strong failures of serial independence (e.g., long-range memory as in 1/f "pink" noise) dissociates the scaling properties from any ergodic Gaussian properties (Box 4; Fig. 1)[6,68,69].

***

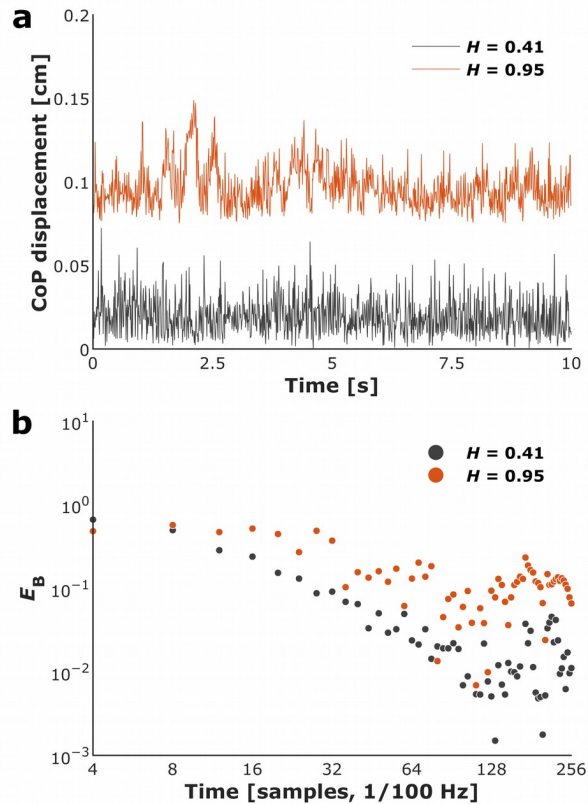

**Fig. 1.** Nonlinear intermittency in physiological signals is nonergodic. (**a**) Two representative trajectories of healthy human center of pressure (CoP) displacement, one showing white Gaussian noise (Hurst exponent, $H$ = 0.41) and the other showing long-range temporal correlations or fractality ($H$ = 0.95). Note that a signal shows fractality if 0.5 < $H$ < 1; the closer $H$ is to 1, the stronger the fractality. (**b**) The ergodicity breaking parameter, $E_B$ converges to zero in the limit $t \rightarrow \infty$ if the signal contains white Gaussian noise but may not converge at all if the signal shows fractality. A fuller treatment of the relationship between $E_B$ and Hurst exponent has been presented elsewhere[6].

\*\*\*

This latter point suggests something rather broad about the potential future of negotiating a constructive rapport between our ergodic traditions and our growing grasp of nonergodicity. Here, we see an essential bridge to build between the academic's luxury to delve into the fine-print entailment of ergodicity and the clinician's urgency for practical use, that is, between the mathematical curiosity and the need for a clear interpretation. Specifically, the ergodic models that adjudicate the unique effect sizes and significance testing are ready to incorporate nonergodic aspects of our measurements. The challenge is to identify those ergodic estimates of nonergodicity, which require taking the big-picture appreciation that, perhaps, what is nonergodic at the fine-grain of our measurements submits to coarser-grained nonlinear-dynamical quantification. Once ergodically described, as in the case of multifractal estimates, nonergodicity can swim amidst the same statistical waters as the classical ergodic measures that clinicians might be more comfortable with[70,71]. For instance, VAR models can test mutual effects amongst multifractal estimates from the same system[67,68]. So at the minimum, ergodic estimates of nonergodicity might be submitted to the same models that clinicians need to identify valid biomarkers.

An immensely useful possibility is that clinically most relevant measures of nonergodicity might align with the greatest ergodicity. For instance, fractal estimates of multifractal, intermittent processes can hover around a mean. While maintaining all statistical evidence of nonlinear intermittency, prior changes in fractal scaling might predict later corrective effects, hence producing significantly narrower multifractal variety. This outcome appears to correlate clearly with a stable reduction of postural sway[58,72]. That is, some of these measures of nonergodic phenomena can behave ergodically.

## Conclusions

Understanding biological and psychological behavior requires a broad range of approaches and methods and faces the fundamental challenge of deciphering the "choreography" associated with complex behaviors and functions. It involves making "sense" of vast amounts of data of many types at multiple scales in time and space. Basing such a scientific project upon false assumptions about the nature of the data might be catastrophic. We have made a case that part of the reproducibility crisis[8–17], and the slow growth of knowledge about nonlinear, far-from-equilibrium dynamics, can be attributed to the violations of ergodic assumptions. We have offered the possibility that both these expectations could immensely benefit from respecting nonergodicity. The stochastic process involved in the study phenomena needs to be made explicit, and the nonergodic data needs to be transformed to find an appropriate ergodic observable. This new observable can be modeled to obtain stable statistical outcomes. Taking such steps will lead to a paradigm shift, allowing researchers to investigate the nonstationary, far-from-equilibrium processes that characterize the creativity and emergence of biological and psychological behavior.

**ACKNOWLEDGMENTS**

We thank Ole Peters for valuable suggestions that helped solidify some of the ideas presented in this review.

**COMPETING FINANCIAL INTERESTS**

The authors declare no competing financial interests.

**REFERENCES**

1.   Hamaker, E. Why researchers should think "within-person": A paradigmatic rationale. in *Handbook of Research Methods for Studying Daily Life* 43–61 (The Guilford Press, 2012).

2.   Kievit, R., Frankenhuis, W., Waldorp, L. & Borsboom, D. Simpson's paradox in psychological science: A practical guide. *Front. Psychol.* **4**, 513 (2013).

3.   Fisher, A. J., Medaglia, J. D. & Jeronimus, B. F. Lack of group-to-individual generalizability is a threat to human subjects research. *Proc. Natl. Acad. Sci.* **115**, E6106–E6115 (2018).

4.   Adolf, J., Schuurman, N. K., Borkenau, P., Borsboom, D. & Dolan, C. V. Measurement invariance within and between individuals: A distinct problem in testing the equivalence of intra- and inter-individual model structures. *Front. Psychol.* **5**, 883 (2014).

5.   Thirumalai, D., Mountain, R. D. & Kirkpatrick, T. R. Ergodic behavior in supercooled liquids and in glasses. *Phys. Rev. A* **39**, 3563–3574 (1989).

6.   Deng, W. & Barkai, E. Ergodic properties of fractional Brownian-Langevin motion. *Phys. Rev. E* **79**, 11112 (2009).

7.   Berman, Y., Peters, O. & Adamou, A. Wealth inequality and the ergodic hpothesis: Evidence from the United States. *Available SSRN 2794830* (2019).

8.   Prinz, F., Schlange, T. & Asadullah, K. Believe it or not: How much can we rely on published data on potential drug targets? *Nat. Rev. Drug Discov.* **10**, 712 (2011).

9.   Van Bavel, J. J., Mende-Siedlecki, P., Brady, W. J. & Reinero, D. A. Contextual sensitivity in scientific reproducibility. *Proc. Natl. Acad. Sci.* **113**, 6454–6459 (2016).

10.  Fang, F. C. & Casadevall, A. Retracted science and the retraction index. *Infect. Immun.* **79**, 3855–3859 (2011).

11.  Mobley, A., Linder, S. K., Braeuer, R., Ellis, L. M. & Zwelling, L. A survey on data reproducibility in cancer research provides insights into our limited ability to translate findings from the laboratory to the clinic. *PLoS One* **8**, e63221 (2013).

12. Estimating the reproducibility of psychological science. *Science (80-. ).* **349**, aac4716 (2015).

13. Gilbert, D. T., King, G., Pettigrew, S. & Wilson, T. D. Comment on "Estimating the reproducibility of psychological science". *Science (80-. ).* **351**, 1037–1037 (2016).

14. Botvinik-Nezer, R. *et al.* Variability in the analysis of a single neuroimaging dataset by many teams. *Nature* **582**, 84–88 (2020).

15. Eklund, A., Nichols, T. E. & Knutsson, H. Cluster failure: Why fMRI inferences for spatial extent have inflated false-positive rates. *Proc. Natl. Acad. Sci.* **113**, 7900–7905 (2016).

16. Kriegeskorte, N., Simmons, W. K., Bellgowan, P. S. F. & Baker, C. I. Circular analysis in systems neuroscience: The dangers of double dipping. *Nat. Neurosci.* **12**, 535–540 (2009).

17. Stikov, N., Trzasko, J. D. & Bernstein, M. A. Reproducibility and the future of MRI research. *Magn. Reson. Med.* **82**, 1981–1983 (2019).

18. Wallach, J. D., Boyack, K. W. & Ioannidis, J. P. A. Reproducible research practices, transparency, and open access data in the biomedical literature, 2015–2017. *PLOS Biol.* **16**, e2006930 (2018).

19. Hardwicke, T. E. *et al.* Data availability, reusability, and analytic reproducibility: Evaluating the impact of a mandatory open data policy at the journal Cognition. *R. Soc. Open Sci.* **5**, 180448 (2020).

20. Gilmore, R. O., Diaz, M. T., Wyble, B. A. & Yarkoni, T. Progress toward openness, transparency, and reproducibility in cognitive neuroscience. *Ann. N. Y. Acad. Sci.* **1396**, 5–18 (2017).

21. Halsey, L. G., Curran-Everett, D., Vowler, S. L. & Drummond, G. B. The fickle P value generates irreproducible results. *Nat. Methods* **12**, 179–185 (2015).

22. Halsey, L. G. The reign of the *p*-value is over: What alternative analyses could we employ to fill the power vacuum? *Biol. Lett.* **15**, 20190174 (2019).

23. Button, K. S. *et al.* Power failure: Why small sample size undermines the reliability of neuroscience. *Nat. Rev. Neurosci.* **14**, 365–376 (2013).

24. Maxwell, S. E., Kelley, K. & Rausch, J. R. Sample size planning for statistical power and accuracy in parameter estimation. *Annu. Rev. Psychol.* **59**, 537–563 (2007).

25. Krzywinski, M. & Altman, N. Error bars. *Nat. Methods* **10**, 921–922 (2013).

26. Ho, J., Tumkaya, T., Aryal, S., Choi, H. & Claridge-Chang, A. Moving beyond P values: data analysis with estimation graphics. *Nat. Methods* **16**, 565–566 (2019).

27. Ioannidis, J. P. A., Munafò, M. R., Fusar-Poli, P., Nosek, B. A. & David, S. P. Publication and other reporting biases in cognitive sciences: Detection, prevalence, and prevention. *Trends Cogn. Sci.* **18**, 235–241 (2014).

28. Molenaar, P. C. M. A manifesto on psychology as idiographic science: Bringing the person back Into scientific psychology, this time forever. *Meas. Interdiscip. Res. Perspect.* **2**, 201–218 (2004).

29. Molenaar, P. C. M. On the implications of the classical ergodic theorems: Analysis of developmental processes has to focus on intra-individual variation. *Dev. Psychobiol.* **50**, 60–69 (2008).

30. Lowie, W. M. & Verspoor, M. H. Individual differences and the ergodicity problem. *Lang. Learn.* **69**, 184–206 (2019).

31. Medaglia, J. D., Ramanathan, D. M., Venkatesan, U. M. & Hillary, F. G. The challenge of non-ergodicity in network neuroscience. *Netw. Comput. Neural Syst.* **22**, 148–153 (2011).

32. Molenaar, P. C. M., Sinclair, K. O., Rovine, M. J., Ram, N. & Corneal, S. E. Analyzing developmental processes on an individual level using nonstationary time series modeling. *Dev. Psychol.* **45**, 260–271 (2009).

33. Molenaar, P. C. M. & Campbell, C. G. The new person-specific paradigm in psychology. *Curr. Dir. Psychol. Sci.* **18**, 112–117 (2009).

34. Fitelson, B. Confirmation, causation, and Simpson's paradox. *Episteme* **14**, 297–309 (2017).

35. Lerman, K. Computational social scientist beware: Simpson's paradox in behavioral data. *J. Comput. Soc. Sci.* **1**, 49–58 (2018).

36. Bains, W. What do we think life is? A simple illustration and its consequences. *Int. J. Astrobiol.* **13**, 101–111 (2014).

37. Hasselman, F. & Bosman, A. M. T. Studying complex adaptive systems with internal states: A recurrence network approach to the analysis of multivariate time-series data representing self-reports of human experience. *Front. Appl. Math. Stat.* **6**, 9 (2020).

38. Hamaker, E. L., Dolan, C. V & Molenaar, P. C. M. Statistical modeling of the individual: Rationale and application of multivariate stationary time series analysis. *Multivariate Behav. Res.* **40**, 207–233 (2005).

39. Castro-Schilo, L. & Ferrer, E. Comparison of nomothetic versus idiographic-oriented methods for making predictions about distal outcomes from time series data. *Multivariate Behav. Res.* **48**, 175–207 (2013).

40. Kulkarni, A. M., Dixit, N. M. & Zukoski, C. F. Ergodic and non-ergodic phase transitions in globular protein suspensions. *Faraday Discuss.* **123**, 37–50 (2003).

41. Manzo, C. *et al.* Weak ergodicity breaking of receptor motion in living cells stemming from random diffusivity. *Phys. Rev. X* **5**, 11021 (2015).

42. Nosonovsky, M. & Roy, P. Allometric scaling law and ergodicity breaking in the vascular system. *Microfluid. Nanofluidics* **24**, 53 (2020).

43. Fitts, P. M. The information capacity of the human motor system in controlling the amplitude of movement. *J. Exp. Psychol.* **47**, 381–391 (1954).

44. Krakauer, J. W., Hadjiosif, A. M., Xu, J., Wong, A. L. & Haith, A. M. Motor Learning. in *Comprehensive Physiology* 613–663 (2019). doi:doi:10.1002/cphy.c170043.

45. Dutilh, G., Wagenmakers, E.-J., Visser, I. & van der Maas, H. L. J. A phase transition model for the speed-accuracy trade-off in response time experiments. *Cogn. Sci.* **35**, 211–250 (2011).

46. Pacheco, M. M., Lafe, C. W. & Newell, K. M. Search strategies in the perceptual-motor workspace and the acquisition of coordination, control, and skill. *Front. Psychol.* **10**, 1874 (2019).

47. Siegler, R. S. The perils of averaging data over strategies: An example from children's addition. *J. Exp. Psychol. Gen.* **116**, 250–264 (1987).

48. Shaw, G. L., Harth, E. & Scheibel, A. B. Cooperativity in brain function: Assemblies of approximately 30 neurons. *Exp. Neurol.* **77**, 324–358 (1982).

49. Vaadia, E. *et al.* Dynamics of neuronal interactions in monkey cortex in relation to behavioural events. *Nature* **373**, 515–518 (1995).

50. Shadlen, M. N. & Newsome, W. T. The variable discharge of cortical neurons: Implications for connectivity, computation, and information coding. *J. Neurosci.* **18**, 3870–3896 (1998).

51. Masuda, N. & Aihara, K. Ergodicity of spike trains: When does trial averaging make sense? *Neural Comput.* **15**, 1341–1372 (2003).

52. Peters, O. & Maximilian, W. A recipe for irreproducible results. *arXiv* 1706.07773v1 (2017).

53. Loetscher, T., Schwarz, U., Schubiger, M. & Brugger, P. Head turns bias the brain's internal random generator. *Curr. Biol.* **18**, R60–R62 (2008).

54. Hilbert, M. Toward a synthesis of cognitive biases: How noisy information processing can bias human decision making. *Psychol. Bull.* **138**, 211–237 (2012).

55. Wallot, S. & Kelty-Stephen, D. G. Interaction-dominant causation in mind and brain, and its implication for questions of generalization and replication. *Minds Mach.* **28**, 353–374 (2018).

56. Van Orden, G. C., Holden, J. G. & Turvey, M. T. Human cognition and $1/f$ scaling. *J. Exp. Psychol. Gen.* **134**, 117–123 (2005).

57. Brown, S. D. & Steyvers, M. Detecting and predicting changes. *Cogn. Psychol.* **58**, 49–67 (2009).

58. Kelty-Stephen, D. G., Lee, I.-C., Carver, N. S., Newell, K. M. & Mangalam, M. Visual effort moderates a self-correcting nonlinear postural control policy. *bioRxiv* 209502 (2020) doi:10.1101/2020.07.17.209502.

59. Mangalam, M., Lee, I.-C., Newell, K. M. & Kelty-Stephen, D. G. Visual effort moderates postural cascade dynamics. *bioRxiv* 209486 (2020) doi:10.1101/2020.07.17.209486.

60. Kuznetsov, N. & Wallot, S. Effects of accuracy feedback on fractal characteristics of time estimation. *Front. Integr. Neurosci.* **5**, 62 (2011).

61. Richardson, L. F. The analogy between mental images and sparks. *Psychol. Rev.* **37**, 214–227 (1930).

62. Hollich, G., Hirsh-Pasek, K. & Golinkoff, R. Breaking the language barrier: An emergentist coalition model for the origins of word learning. in *Monographs for the Society for Research in Child Development* 65 (3, Serial No. 262) (2000).

63. Box, G. E. P. & Pierce, D. A. Distribution of residual autocorrelations in autoregressive-integrated moving average time series models. *J. Am. Stat. Assoc.* **65**, 1509–1526 (1970).

64. Werner, G. Fractals in the nervous system: conceptual implications for theoretical neuroscience. *Front. Physiol.* **1**, 15 (2010).

65. Kilian, L. & Lütkepohl, H. *Structural Vector Autoregressive Analysis*. (Cambridge University Press, 2017).

66. Podobnik, B. & Stanley, H. E. Detrended cross-correlation analysis: A new method for analyzing Two nonstationary time series. *Phys. Rev. Lett.* **100**, 84102 (2008).

67. Mangalam, M., Carver, N. S. & Kelty-Stephen, D. G. Global broadcasting of local fractal fluctuations in a bodywide distributed system supports perception via effortful touch. *Chaos, Solitons & Fractals* **135**, 109740 (2020).

68. Mangalam, M., Carver, N. S. & Kelty-Stephen, D. G. Multifractal signatures of perceptual processing on anatomical sleeves of the human body. *J. R. Soc. Interface* **17**, 20200328 (2020).

69.     Mangalam, M. & Kelty-Stephen, D. G. Multiplicative-cascade dynamics supports whole-body coordination for perception via effortful touch. *Hum. Mov. Sci.* **70**, 102595 (2020).

70.     Peters, O. The ergodicity problem in economics. *Nat. Phys.* **15**, 1216–1221 (2019).

71.     Peters, O. & Gell-Mann, M. Evaluating gambles using dynamics. *Chaos An Interdiscip. J. Nonlinear Sci.* **26**, 23103 (2016).

72.     Kelty-Stephen, D. G. Multifractal evidence of nonlinear interactions stabilizing posture for phasmids in windy conditions: A reanalysis of insect postural-sway data. *PLoS One* **13**, e0202367 (2018).